\documentclass[12pt]{article}

\usepackage{graphicx}
\usepackage{amssymb}

\begin{document}

\title{Bounds on new light particles from high-energy and very small 
momentum transfer $np$ elastic scattering data}
\author{Yuri Kamyshkov\thanks{kamyshkov@utk.edu},~Jeffrey Tithof\thanks{jtithof@utk.edu}, \\
{\it University of Tennessee, TN 37996-1200, USA} \\
Mikhail Vysotsky\thanks{vysotsky@itep.ru} \\
{\it ITEP, Moscow, 117218, Russia}}
\date{}

\maketitle

\baselineskip=11.6pt

\begin{abstract}
\small
We found that spin-one new light particle exchanges are
strongly bounded by high-energy and small momentum transfer 
$np$ elastic scattering data; the analogous bound for a scalar particle 
is considerably weaker, while for a pseudoscalar particle  no bounds 
can be set. These bounds are compared with the bounds extracted 
from low-energy $n-Pb$ scattering experiments and from the 
bounds of $\pi^0$ and $K^+$ meson decays.
\end{abstract}
\normalsize
\linespread{1.25}

\section{Introduction}

The Standard Model of three fundamental forces describes
interactions of elementary particles very well.
While the electromagnetic force has a long interaction range, the 
short radius of the ``weak force" ($\sim$ 1/1000 fermi) is determined 
by the heavy masses of mediating W and Z bosons ($\sim$ 100 GeV). 
The QCD forces are typically contained within a confinement radius 
of $\sim$ 1 fermi. The effects of the long-range gravitational force can usually 
be neglected in elementary particle scattering experiments.

The search for new forces of nature is a major goal of experiments at 
high-energy colliders. Rare transitions and decays of fundamental 
particles can also shed light on new interactions. These 
experiments are probing for new forces at distances shorter than 
1/1000 fermi. However, the existence of new forces at distances 
larger than the confinement radius of a nucleon ($\sim$ 1 fermi) can 
also be probed and constrained by sensitive experiments. 

Experimental searches for and limits on these new forces of nature can be  
pursued in two directions: (a) as deviations from the Newtonian 
law of gravity where the new force is expressed as a modification 
of the $1/r^2$ law, usually by an additional Yukawa term that can be parameterized 
with two parameters $\alpha$, the relative strength of new interaction, 
and $\lambda$, the characteristic radius of the interaction. 
Applied in the analysis of experimental data at macroscopic distances 
down to $\sim$ a micrometer, this ansatz describes the possible 
deviations from classical gravity; (b) as a quantum field theory  
description of the interactions (excluding gravity) in a covariant form, 
which can be expressed in the lowest perturbation order through the 
coupling constant $g$ and the mass of the exchanged particle
mediating the interaction $\mu$. Covariant forms that can be consistently 
considered in this description are scalar (S), pseudoscalar (P), 
vector (V), and axial vector (A). We can argue that higher spins of the 
intermediate particle should not be considered since they lead to
non-renormalizable theory. 
The particles that mediate the new force could be absent from 
the spectrum of known particles \cite{PDG} due to their small 
mass and coupling constant or due to some other reason that 
is helping them avoid detection. In any case, if these particles
are not observed, direct experimental limits on their existence
in terms of $g$ and $\mu$ are required.  

In Section 2 of the present paper, we reanalyze the experimental 
small-angle $np$-elastic scattering data at high energy \cite{na6} 
in terms of  bounds on the existence of new forces expressed as 
S, P, V, or A covariant interactions. In Section 3, we examine 
bounds that can be obtained from lower energy data.

\section{Bounds from high-energy $np$ scattering}

The data for small-angle $np$-elastic scattering at high-energy 
were obtained in the NA-6 experiment 
\cite{na6} performed at CERN SPS a quarter century ago. Incident 
neutron energy in the experiment was 100--400 GeV, while the square 
of the 4-momentum transfer $|t|$ was varied in the range $6 \cdot 10^{-6}$ 
to $5 \cdot 10^{-1}$ GeV$^2$. The data of this experiment are consistent 
with extrapolation of the hadronic amplitude from higher $|t|$ values, 
while at $|t| < 10^{-4}$ GeV$^2$ the differential cross-section rises
due to Schwinger scattering, which is the interaction of the neutron's 
magnetic moment with the Coulomb field of the proton or electron. 
The purpose of NA-6 \cite{na6} was to measure hadronic interactions 
at high $s$ in the region of momentum transfer ($\sim |t|<10^{-2}$ 
GeV$^2$) that was usually inaccessible in the scattering of charged 
hadrons due to Coulomb interactions. This is the region where the 
effect of a new force mediated by a light particle may be present.  

Figure \ref{fig1} (similar to Fig. 16 from \cite{na6}) demonstrates that $np$
elastic scattering data in this experiment are well described by 
the following formula:

\begin{figure}[htbp]
\centering
\includegraphics[width=5.0in, keepaspectratio]{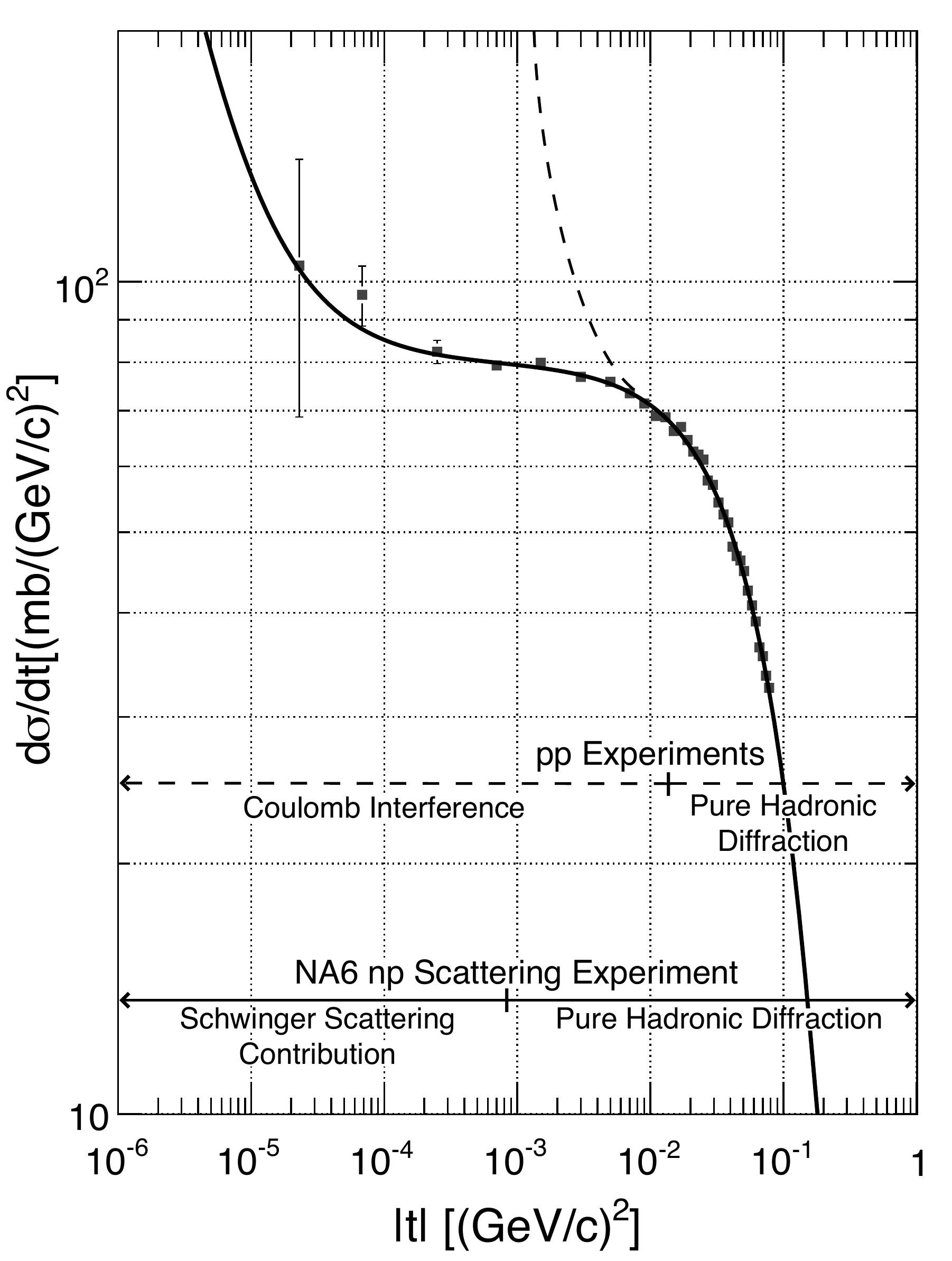}
\caption{\label{fig1} \small Elastic differential 
neutron-proton cross-sections measured in experiment \cite{na6}.
For comparison, the $|t|$ region measurable in $pp$ scattering is 
shown with the effect of the Coulomb interaction indicated by the dashed line.}
\end{figure}

\begin{equation}
\frac{d\sigma}{dt} = A {\rm exp}[bt] - 2 \left(\frac{\alpha
k_n}{m_n}\right)^2 \frac{\pi}{t} \;\; , \label{1}
\end{equation} where $A = (79.78 \pm 0.26) mb/{\rm GeV}^2$ and 
$b = (11.63 \pm 0.08)$
GeV$^{-2}$ were determined from the fit to the data 
(data are taken from Table 7 in \cite{na6}), $m_n$ is the neutron 
mass, and $k_n = -1.91$ is the neutron magnetic moment in 
nuclear magnetons. The factor of 2 in the Schwinger term, as will be
discussed later, accounts for the scattering of the neutron's magnetic 
moment on the proton plus an incoherent contribution of scattering 
on electrons (gaseous hydrogen was used in \cite{na6} as a target). 
Smaller effects  due to neutron polarizability are not included in
the description of the data. This description (\ref{1}) works rather 
satisfactorily with $\chi^2$=41.5 for 31 degrees of freedom.  We will
refer to this description as the ``zero model" since no new force contributions 
are included here.

Although Quantum Chromodynamics does not provide a detailed 
theoretical description of the hadronic elastic scattering at small
$|t|$, i.e. at large impact parameters, hadronic scattering 
has been studied experimentally in great detail in the past and was 
phenomenologically well understood, e.g. in the framework of 
Regge models. The description of elastic data by a single exponent 
was a general universal feature of hadronic scattering observed 
at low $|t|$ in the region where it was not obstructed by Coulomb 
scattering (for example, see \cite{elasticreview} and also the comparison 
with other experiments in \cite{na6}). This justifies, at a phenomenological 
level, our choice of the hadronic scattering description with a single exponent.
However, in an attempt to improve the description of 
the data \cite{na6}, we have tried several alternative modifications of the
exponential term in the ``zero model" (\ref{1}) involving additional 
parameters, including a quadratic term in the exponent  
and the sum of two exponents. In all of these cases, 
$\chi^2$ per degrees of freedom was slightly increased demonstrating 
that more complicated modifications of the ``zero model" are not  
statistically justifiable.

We describe the contribution of a new interaction in the following 
way: Let us suppose that a new light particle with mass $\mu$ exists
which interacts with the neutron and proton with couplings $g_n$
and $g_p$ correspondingly. Assuming scalar, pseudoscalar,
vector, and axial vector couplings of this particle with nucleons,
we obtain the following addition to expression (\ref{1}):

\begin{equation}
\frac{d\sigma_i}{dt} (g,\mu)|_{\rm new} = \frac{|A_i|^2 \cdot FF}
{16\pi s (s-4m^2)} \;\; , \label{2}
\end{equation} where $s = (p_n + p_p)^2$ is the invariant energy 
square and $m$ is the nucleon mass. We parameterize the hadronic 
form factor's contribution as:

\begin{equation}
FF = \frac{1}{(1-t/\Lambda^2)^8} \; \;\; , \label{3}
\end{equation} which comes from a $1/q^4$ decrease of the 
nucleon form factor, and we set $\Lambda$ equal to the mass of 
the lightest meson resonance with appropriate quantum 
numbers ($\eta'$ in the case of pseudoscalar). Finally,
we use the following amplitude squares for different couplings:

\begin{equation}
|A_S|^2 = \frac{g_S^4}{(t-\mu^2)^2}(4m^2 -t)^2 \;\; , \label{4}
\end{equation}

\begin{equation}
|A_P|^2 = \frac{g_P^4 t^2}{(t-\mu^2)^2} \;\; , \label{5}
\end{equation}

\begin{equation}
|A_V|^2 = \frac{4g_V^4}{(t-\mu^2)^2} [s^2 - 4m^2 s + 4m^4 + st +
\frac{1}{2} t^2] \;\; , \label{6}
\end{equation}

\begin{equation}
|A_A|^2 = \frac{4g_A^4}{(t-\mu^2)^2} [s^2 + 4m^2 s + 4m^4 +st +
\frac{1}{2} t^2 +\frac{4m^4 t^2}{\mu^4} +\frac{8m^4 t}{\mu^2} ] \;\; , \label{7}
\end{equation} where coupling constants $g_i^2 \equiv g_p^i g_n^i$.
\vskip 0.2cm

It is quite natural to suppose that a new light particle's couplings
with nucleons originates from its couplings with quarks.
In this case, (6) and (7) are modified. For the vector exchange, the induced
magnetic moment's interaction term should be added to the
scattering amplitude. Since its numerator contains momentum
transfer divided by $m_N$, which in considered kinematics
gives a factor much smaller than 1, we can safely neglect it and 
use (6) in what follows. The case of the axial vector exchange is more 
delicate and discussed in detail in the Appendix.

We can now turn to the discussion of other features of  the
$np$ elastic scattering amplitude. Though the strong interaction 
amplitude cannot be determined theoretically from the first 
principles, our confidence that $1/t$ dependence is absent 
in strong interactions for 
$|t|<m^2_{\pi}$ opens the road to bounding the light particle exchange
if its mass is smaller than that of the $\pi$-meson. Experimental data at
$|t|<m^2_{\pi}$ matter for our bounds, which makes the precise value of 
$\Lambda$ in the expression for $FF$ not important, since in
the relevant domain of $|t|$ the form factor is close to $1$.
For the same reason, no form factor is introduced for the Schwinger 
term in (\ref{1}).

For each fixed set of parameters $g_{i}^2$ and $\mu$ 
describing the possible contribution of a ``new force", we are 
fitting the experimental distribution with a combined function
(\ref{1})+(\ref{2}), where parameters $A$ and $b$ describing 
the standard hadronic contribution are free. Then the maps of $A$, $b$,
and the minimum values of $\chi^2$ are composed as 
functions of $g_{i}^2$ and $\mu$. Analyzing the $\chi^2$ map, we determined 
the level of $\chi^2$ \cite{stat} above which parameters of the
``new force" become incompatible with experimental data at a
confidence level (C.L.) greater than 90\%. At this level we also examined and ensured 
that parameters $A$ and $b$ remain within the 90\% C.L. close to 
those in the ``zero model". In this way, we can ensure that the 
``new force" contribution does not substitute for the standard hadronic 
plus electromagnetic contributions in the description of the data.

\begin{figure}[htbp]
\centering
\includegraphics[width=4.5in,keepaspectratio]{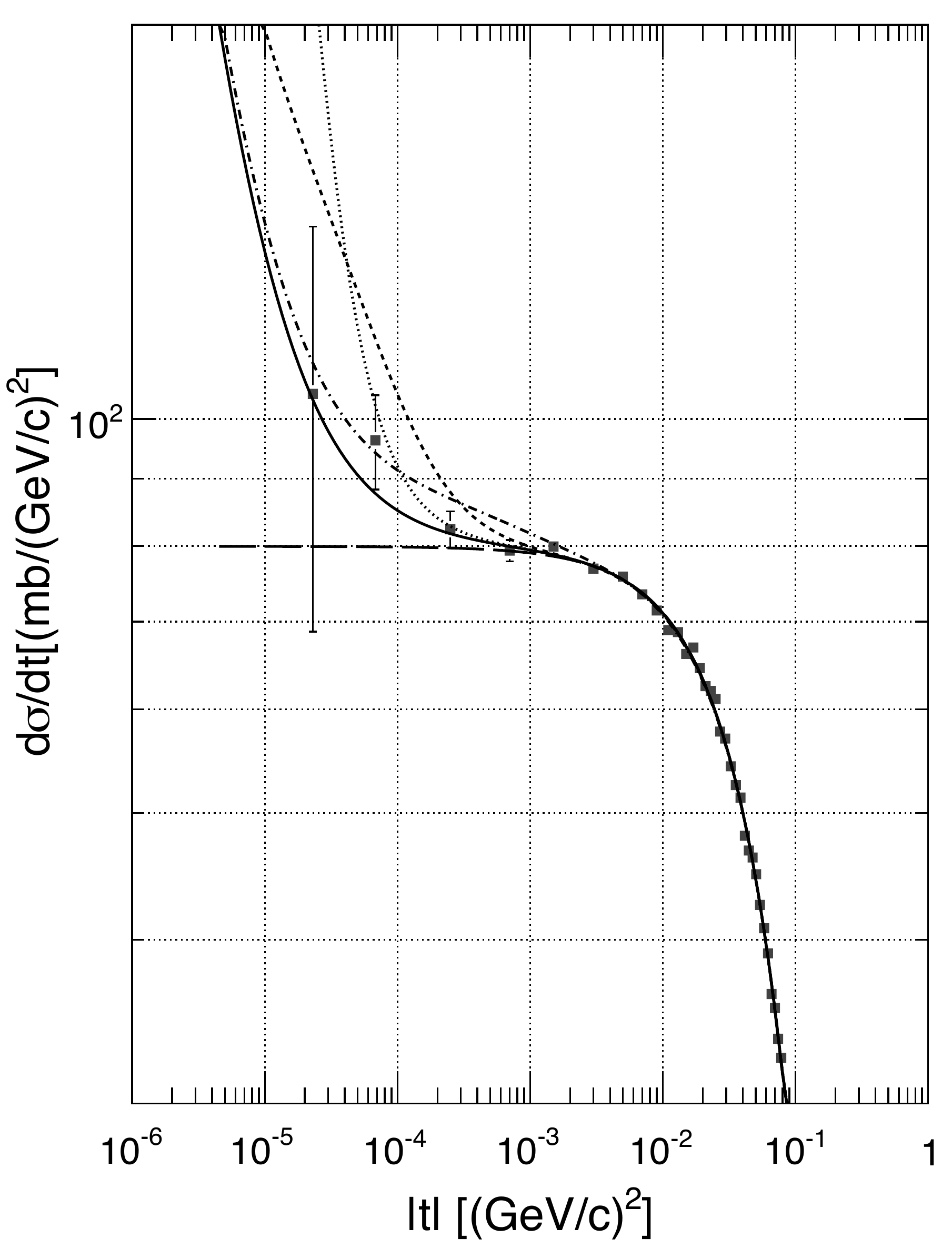}
\caption{\label{fig2} \small Several fits to the experimental data
of \cite{na6}: long-dash line -- single exponent without the Schwinger 
contribution; solid line -- the ``zero model" description of (\ref{1});
dotted line --  the ``zero model" plus the new vector particle contribution
with $\mu$ = 1 MeV and $g^2=0.0015$;
short-dash line --  the ``zero model" plus the new vector particle contribution
with $\mu$ = 10 MeV and $g^2=0.005$;
dot-dashed line --  the ``zero model" plus the new vector particle contribution
with $\mu$ = 40 MeV and $g^2=0.025$.}

\end{figure}

Figure \ref{fig2} shows, for comparison, fits to the data for the ``zero 
model" and for several excluded models for the new vector particle 
exchange with parameters slightly beyond the excluded limits for  
$\mu_V$ and $g_V^2$.
 
Two comments need to be made on formulas (\ref{4})--(\ref{7}): 
(a) the amplitudes with the exchange in the $t$-channel of a point 
like particle with spin $\alpha$ depend on $s$ as $s^\alpha$. That 
results in an amplitude behavior of $s^0$ for scalar and pseudoscalar 
and of $s$ for vector and axial vector particles. This property of 
high-energy scattering amplitudes would allow us to determine the 
value of the spin of the ``new physics" mediator; (b) the pseudoscalar exchange 
vanishes at $t=0$. 
These comments explain why we will get the strongest bounds on 
$g_A$ and $g_V$, a weaker bound on $g_S$, and no bound on $g_P$.

Before presenting fit results, let us explain why we neglect the
interference of a new particle exchange amplitude with the strong
amplitude and with the photon exchange (Schwinger) amplitude.
The strong amplitude is almost entirely imaginary in the energy
domain studied in \cite{na6} ($|Re/Im|<0.1$ \cite{PDG}). That 
is why it does not interfere with the real amplitude of a point-like new 
particle exchange. Interference of the strong amplitude (as well as
the new force amplitude) with the Schwinger term is negligible since the 
interference term is constant at $t=0$  \footnote{The Schwinger part 
of the interference term contains $q_\mu/q^2$, $q^2 \equiv t$, 
which multiplies $q_\mu \equiv (p_1 - p_2)_\mu$ cancelling 
the $1/q^2$ enhancement, or  $(p_1 + p_2)_\mu$ giving zero.}, 
unlike the square of the Schwinger amplitude, which
contributes significantly at small $|t|$ because of $1/t$ behavior. 

\begin{figure}[htbp]
\begin{center}
\includegraphics[width=5.3in,keepaspectratio]{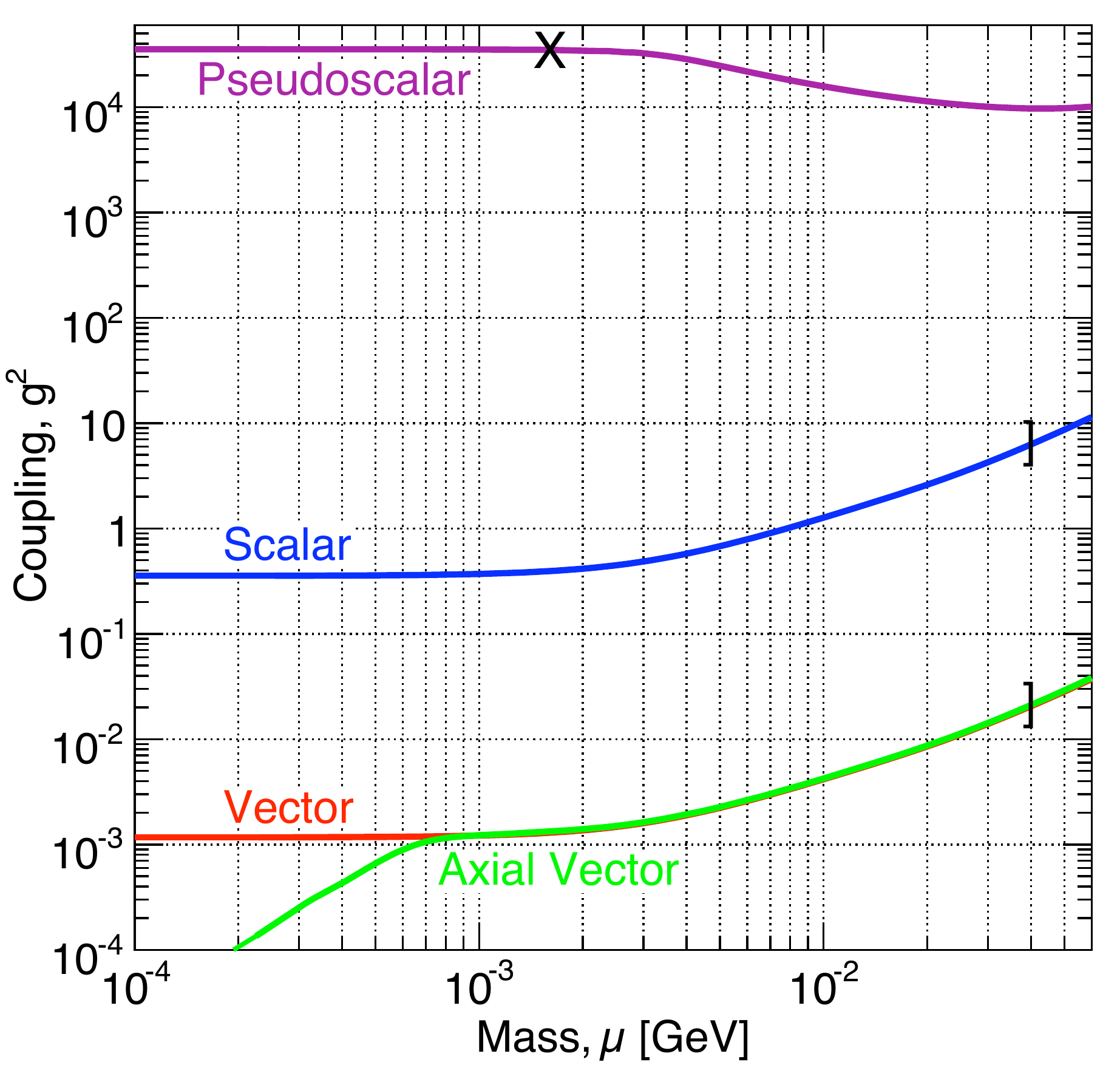}
\caption{\label{fig3} \small Compilation of the upper bounds obtained in the 
current analysis in terms of $g^2$ and $\mu$ at a 90\% C.L. No limit can be 
set for the pseudoscalar exchange. Brackets indicate the interval of 
mass $\mu$ where the analysis was validated (see text).}
\end{center} 
\end{figure}

The mass of a light particle $\mu$ was bounded 
in our fits to be below 100 MeV, and the range of the coupling constants 
varied depending on the particular model. Compilation of the bounds 
obtained from the $\chi^2$ limit for P, S, V, and A models in coordinates 
$g^2$ versus $\mu$ is presented in Figure \ref{fig3}.

In the next step of analysis, we checked that for each model, fitted 
parameters $A$ and $b$ corresponding to the boundary of the excluded 
domain of $g_i^2$ and $\mu$ must not deviate from their ``zero model" 
values by more than $1.28\sigma$, where $\sigma$ is the corresponding 
error of parameters $A$ and $b$ from the ``zero model." We found that 
these conditions are satisfied if $\mu < 40 MeV$ for the S, V, and A-models, 
and cannot be satisfied for any value of $\mu$ for the P-model. Thus, our 
bounds on the coupling strength $g^2$ shown in Figure \ref{fig3} should 
only be referred to in these validated domains (indicated in Figure \ref{fig3} 
by brackets). No consistent limit can be set for the P-model. In addition, we
should notice that the very high value of $g_P$ obtained from the $\chi^2$ 
analysis for the P-model makes our perturbative approach of formula (\ref{5})
not valid. We should conclude therefore that the experimental data \cite{na6}
do not provide any limit for the pseudoscalar exchange.

The factor of 2 in the Schwinger term of the ``zero model" in (\ref{1})
is coming from both $n-p$ and $n-e$ scattering and is an estimate of 
equal contribution from both. However, $n-e$ scattering occurs in 
a different kinematical range and the event selection criteria 
in \cite{na6} could suppress the detection of electrons. Consequently, we 
varied the factor in the Schwinger term of the ``zero model" (\ref{1})
from 1 (no $n-e$ contribution) to 3 (double $n-e$ contribution) 
and found in analysis that this variation was not very significant,
changing our limiting value for $g^2$ by $\pm$ 8\% for a fixed 
value of $\mu$. 

One can notice that since the average
$s \approx 540$ GeV$^2$ in experiment \cite{na6}, vector $(A_V)$ and axial vector
$(A_A)$ amplitudes, as follows from Eq. (\ref{6}) and (\ref{7}), are
practically the same (see Figure \ref{fig3}), except when 
$\mu \lesssim 1 MeV$.  In this case, the last two terms of Eq. (\ref{7})
arising from the $q_{\mu}q_{\nu}/{\mu^2}$ part of the propagator
of the axial vector particle start to dominate the amplitude.

Our bounds on the parameters $g_V^2$ and $g_A^2$ (Figure \ref{fig3}) are 
rather strong; say, for $\mu = 10$ MeV, $g_{V,A}^2 < 5 \cdot 10^{-3}$ 
at 90\% C.L., which corresponds to
 
\begin{equation}
g_N^{V,A} < 0.071 \;\; , \label{8}
\end{equation} four times smaller
than the QED coupling constant $\sqrt{4\pi\alpha} \simeq 0.3$.
For the scalar exchange, taking $\mu = 10$ MeV, we get a much weaker
bound, $g_S^2 < 1.4$.

\section {Bounds from lower energy data}

We will now compare our results from the previous section with other
searches for new interactions \cite{6}--\cite{bordag} 
in which new  light particles
participate. In the literature, the effect of new forces
is usually parameterized as a deviation from the Newtonian 
gravitational potential:
\begin{equation}
V(r) = -G_N \frac{m_1 m_2}{r} [1+\alpha_G \exp(-r/\lambda)] \;\; ,
\label{9}
\end{equation} which is an adequate approximation for the 
description of the effect of a new
particle exchange between nonrelativistic constituents. 
The following relationship exists between the coupling
constant $\alpha_G$ and characteristic length $\lambda$ and our
parameters $g_i^2$ and $\mu$ in cases of vector and scalar  
exchanges:
\begin{equation}
\alpha_G = \frac{g_{V,S}^2}{4\pi G_N m_p m_n} = 1.35 \cdot 10^{37}
g_{V,S}^2 \; , \;\; \lg\alpha = \lg g_{V,S}^2 + 37.13 \;\; , \label{10}
\end{equation} 
\begin{equation}
\lambda({\rm cm}) = \frac{1}{\mu ({\rm MeV}) 5.05 \cdot 10^{10}}
\; , \;\; \lg\lambda = -\lg\mu - 10.7 \;\; . \label{11}
\end{equation}

The pseudoscalar exchange would not modify the potential in a nonrelativistic
approximation, while axial coupling leads to an interaction
among spins of constituents.

(A) References \cite{6}--\cite{lamoreaux} show e.g. that values of 
$\alpha_G$ larger than $1$ are excluded for $\lambda$ larger 
than 0.1 mm, while for smaller $\lambda$ the upper bounds on 
$\alpha_G$ rapidly grow, reaching $10^9$ at the micron scale. 

\begin{figure}[htbp]
\begin{center}
\includegraphics[width=5.3in,keepaspectratio]{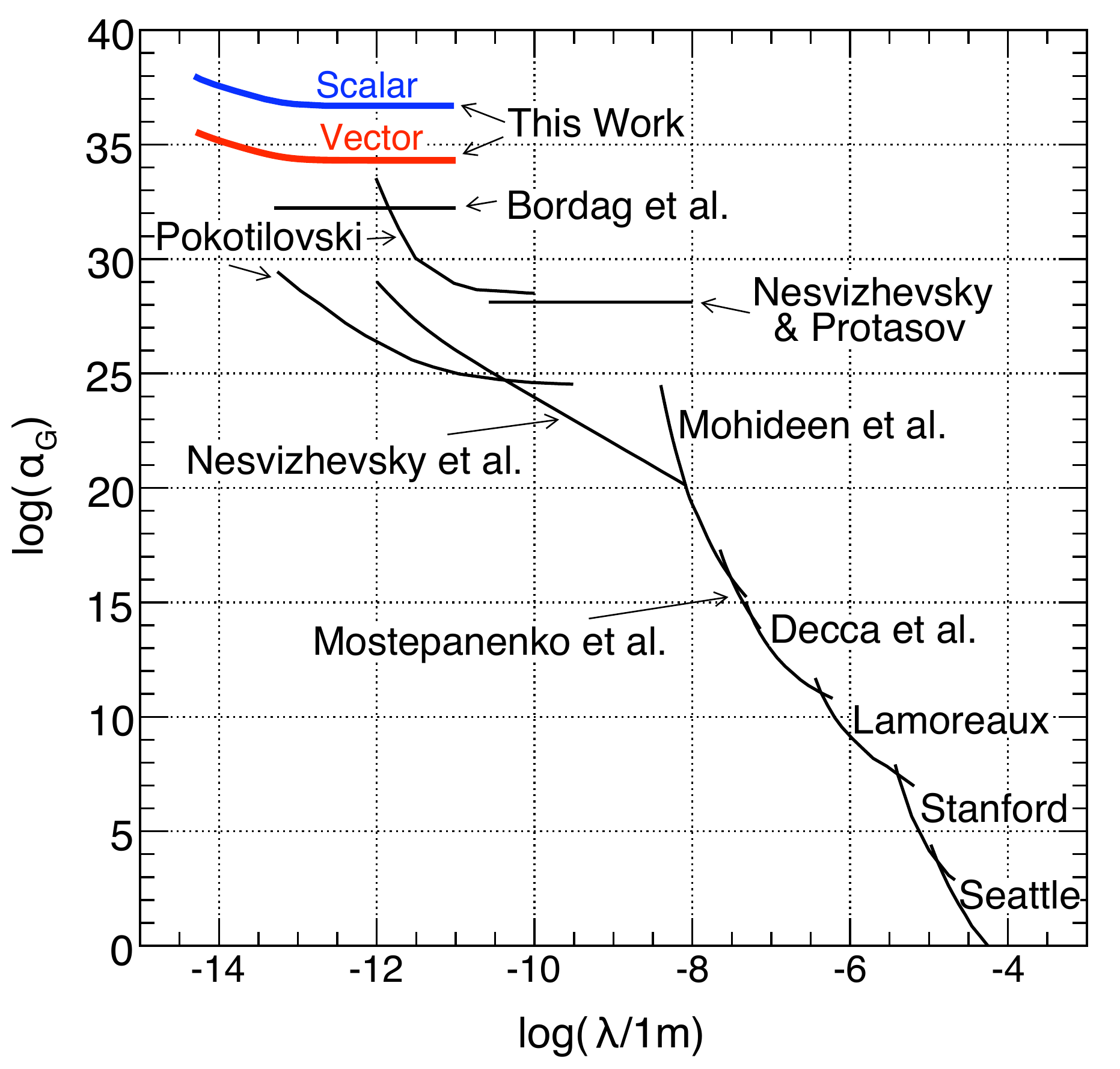}
\end{center}
\caption{\label{fig4} \small Experimental limits on $\alpha_G$ and 
$\lambda$ from \cite{6}--\cite{bordag} parameterizing deviations 
from Newton's law. Our limits transformed into coordinates 
$\alpha_G$ and $\lambda$ are also shown for comparison.}
\end{figure}

In papers \cite{decca}--\cite{bordag}, analogous bounds for shorter 
distances are presented. We see e.g. that for $\lambda = 10^{-13}$ m, 
$\alpha_G$ should be less than $10^{30}$, or $g_{V,S}^2$ less than 
$10^{-7}$. The corresponding value of $\mu$ is 2 MeV. In Figure 
\ref{fig4}, our limits for the V and S particle exchanges in terms 
of $\alpha_G$ and $\lambda$ parameters are compared with the 
limits obtained in papers \cite{6}--\cite{bordag}.

(B) Additionally, the data on low-energy ($1 \; {\rm keV} < E_n < 10$ keV)
neutron scattering on$~^{208}{\rm Pb}$ \cite{3} were applied in
paper \cite{4} to obtain bounds on the possible contributions of
a light scalar particle exchange to neutron-nucleus potential. The upper
bound on the coupling constant of a 10 MeV boson to a nucleon obtained
in \cite{4} corresponds to $g_{V,S}^2 < 4 \cdot 10^{-6}$. This rather 
restrictive bound was obtained from the analysis of the shape of the 
differential cross-section of low-energy $n-~^{208}{\rm Pb}$ scattering, 
where the additional term, originating from the light scalar boson exchange,
leads to a modification of angular dependence not observed in the 
experimental data. Let us stress that the same bound is valid for $g_{V}$.

According to \cite{3,4}, experimental data in the keV energy range
are well described by the following expression:

\begin{equation}
\frac{d\sigma}{d\Omega} = \frac{\sigma_0}{4\pi} [1+\omega E 
\cos\theta] \;\; , \label{12}
\end{equation} where $\sqrt{\sigma_0/4\pi} \approx 10$ fm, and
$\omega = (1.91 \pm 0.42) 10^{-3}$ keV$^{-1}$. These numerical 
values are very reasonable from the nuclear scattering point of 
view and, from the demand that these values are not spoiled by 
a Yukawa potential contribution originating from a light boson 
exchange, bounds on $g$ and $\mu$ were obtained in \cite{4}. 
The point is that the Yukawa
amplitude, interfering with the strong interaction amplitude, will
show up in the following contribution to $\omega$ for $E \to 0$:

\begin{equation}
|\Delta\omega | = \frac{16m_n^2}{\sqrt{\sigma_0/4\pi}}
\frac{\lambda_n^2}{4\pi}\frac{A}{\mu^4} \;\; , \label{13}
\end{equation} and from the demand that $\Delta\omega < \omega$, the above 
mentioned bound was extracted. For an update of results
obtained in \cite{4}, see \cite{5}.

(C) It is quite natural to assume that the coupling of a new light
boson with nucleons originates from its coupling with $u$- and
$d$-quarks. In this case, bounds from pion and kaon decays
\cite{fayet} are applicable. Let us start with vector coupling.
According to CVC, couplings to nucleons are equal to the sum of
the couplings to quarks: $2 f_{uV} + f_{dV}$ for a proton and
$f_{uV} + 2 f_{dV}$ for a neutron ($f_i$ are analogous to our
$g_N^i$). The $\pi^0 \to VV$ decay contributes to $\pi^0 \to
invisible$ decays and, using the experimental bound
$Br(\pi^0\to\nu\nu) < 2.7 \cdot 10^{-7}$ in \cite{fayet}, the following
bound was obtained: 

\begin{equation} 
\sqrt{|f_{uV}^2 - f_{dV}^2|} \leq 4 \cdot 10^{-3} \;\; , 
\label{14} 
\end{equation} 
which is automatically satisfied for an isoscalar 
coupling, $f_{uV} = f_{dV}$. However, the bound on 
the $\pi^0 \to\gamma V$ decay, which contributes into $\pi^0 \to\gamma + 
invisible$ mode, allows bounding of isoscalar couplings as well 
\cite{fayet}: 

\begin{equation} 
\frac{2 f_{uV} + f_{dV}}{3} < 1.6 \cdot 10^{-3} \;\; . 
\label{15} 
\end{equation} 

Here, the experimental bound $Br(\pi^0\to\gamma\nu\nu) < 6 \cdot 
10^{-4}$ was used. These numbers should be compared with our bound 
on $g_N^{V}$ (\ref{8}). 

Since $\pi^0\to SS$ and $\pi^0\to S\gamma$ decays violate 
the corresponding $P$- and $C$-parities, we do not obtain bounds on 
$f_S$ from these decays.  $C$-parity conservation forbids the $\pi^0\to\gamma A$ decay as well, 
while from the bound on $\pi^0\to invisible$ decays, we get the 
coupling constant bound (\ref{14}) for the axial vector boson. 

More stringent upper bounds on the coupling constants follow from 
very strong experimental limits on the branching ratio 
$Br(K^+\to\pi^+ +\nu\nu) < 2\cdot 10^{-10}$.  The longitudinal 
component of the axial vector boson contributes to the decay amplitude 
proportionally as $(2m_q/\mu)f_{qA}$ \cite{fayet}, and even if the axial 
vector boson couples only with light quarks, we obtain: 

\def\la{\mathrel{\mathpalette\fun <}} 
\def\ga{\mathrel{\mathpalette\fun >}} 
\def\fun#1#2{\lower3.6pt\vbox{\baselineskip0pt\lineskip.9pt 
\ialign{$\mathsurround=0pt#1\hfil##\hfil$\crcr#2\crcr\sim\crcr}}}
 
\begin{equation} 
f_{u,d A} \la 10^{-6}\mu ({\rm MeV}) \;\; . 
\label{16} 
\end{equation} 

The factor of $2 m_q/\mu$ is absent when the axial vector interaction is 
substituted by the scalar interaction, and thus we obtain: 

\begin{equation} 
f_{u,d S} \la 10^{-5} \;\; . 
\label{17} 
\end{equation}  

Fortunately, CVC forbids $K\to\pi V$ decays for $\mu^2 = 0$, 
so that is why the bound on the vector coupling for light $\mu$ is not very strong:
 
\begin{equation} 
f_{u,d V} \left(\frac{\mu}{m_K}\right)^2 \la 10^{-5} \;\;. 
\label{18} 
\end{equation}

\section {Conclusions} 

Our bounds obtained from high-energy and very small momentum 
transfer $np$ elastic scattering data \cite{na6} provide exclusions of 
new forces at distances above 5 fermi, which corresponds to 
exchanged particle masses lighter than 40 MeV. These bounds are 
extracted in a covariant approach, as an alternative to the bounds on 
couplings at larger distances, extracted from the absence of deviations 
from the Newtonian gravitational law. 

Both low-energy $n- ^{208}{\rm Pb}$  and high-energy $np$ 
scattering data lead to 
similar upper bounds on the coupling constants  for $\approx 10$ 
MeV vector bosons, though upper bounds from $n$ -- Pb scattering 
on the coupling constant $g_N^{V}$ are $\sim$30 times lower and 
close to the bounds from $\pi^0 \to$ invisible and $\pi^0 
\to\gamma \; +$ invisible decays on the vector coupling constants 
with quarks. 

Strong upper bounds on the ``new physics" contribution into the $K^+\to\pi^+ + 
invisible$ decay allows us to get very strong bounds for scalar and 
axial vector bosons: $g_N^{A,S} \la 10^{-5}$ for a 10 MeV boson mass. 

We are grateful to A. B. Kaidalov, B. Z. Kopeliovich, L.~B. Okun, 
and Yu. N. Pokotilovsky for useful discussions and comments. 
M. V. was partially supported by Rosatom and grants RFBR 07-02-00021, 
RFBR 08-02-00494 and NSh-4568.2008.2. Y.K. would like to 
acknowledge research support from the United States Department 
of Energy HEP grant DE-FG02-91ER40627.

\section {Appendix}

For vanishing light quark masses, their isotriplet axial current is
conserved. The same should hold for the nucleon currents and is
achieved by accounting for pion exchanges:

\begin{eqnarray}
\tilde A_A &=& g_A^2 \bar n \gamma_\beta \gamma_5 n
\left(g_{\alpha \beta} - \frac{k_\alpha k_\beta}{k^2 -
m_\pi^2}\right) \frac{(g_{\alpha\mu} - \frac{k_\alpha
k_\mu}{\mu^2})}{k^2 -\mu^2} \left(g_{\mu\nu} -\frac{k_\mu
k_\nu}{k^2 -m_\pi^2}\right)\bar p \gamma_\nu \gamma_5 p =
\nonumber \\
& = & \frac{g_A^2}{k^2 -\mu^2} \left[g_{\alpha\beta} -
\frac{k_\alpha k_\beta}{\mu^2} \frac{(m_\pi^4 - 2\mu^2 m_\pi^2 +
k^2 \mu^2)}{(k^2 - m_\pi^2)^2}\right] \bar n \gamma_\alpha \gamma_5
n\bar p \gamma_\beta \gamma_5 p \;\; . \label{8'}
\end{eqnarray}

For a massless pion, the $1/\mu^2$ singularity cancels out and the expression
in square brackets contains $k_\alpha k_\beta/k^2$, which, acting
upon fermionic axial currents, becomes $(2m_N)^2/k^2$. The numerator
of the expression for differential cross-section is regular at
$k^2 \equiv t = 0$ since the square of the pseudoscalar exchange amplitude
contains $t^2$ in the numerator (see (5)), while the interference of the axial vector
and pseudoscalar exchanges is proportional to $t$ (the denominator
equals $(t-\mu^2)^2$ independently of the spin of the exchanged
boson).

In real life, light quarks, as well as pions, have nonzero masses, and
to obtain an amplitude square for the axial vector boson exchange we should
substitute $\mu$ by $\tilde\mu$ in the square brackets of (7), where

$$
\frac{1}{\tilde\mu^2} = \frac{1}{\mu^2} \frac{m_\pi^4 - 2\mu^2
m_\pi^2 + t\mu^2}{(t - m_\pi^2)^2} \;\; ,
$$
and for $t, \mu^2 \ll m_\pi^2$ we get $\tilde\mu = \mu$.

The numerator of the expression for the differential cross-section is
singular for $\mu\to 0$. However, in renormalizable theory, the mass of
the axial vector boson equals its gauge coupling constant ($g_A$  in our case)
times the vacuum average of the corresponding higgs field.

As an example, one can have in mind the expansion of the Standard Model
with two higgs doublets with opposite hypercharges, where
Peccei--Quinn $U(1)$-symmetry is spontaneously broken producing an
axion. In order to suppress axion couplings to quarks and leptons, the
additional singlet neutral higgs field $N$ is usually added, which
makes the axion invisible. Gauging of Peccei--Quinn $U(1)$ leads to
the axial coupling of the corresponding vector boson to matter.
Such a light axial vector boson is discussed in particular in \cite{fayet},
where it is light due to the smallness of the gauge coupling constant,
while the vacuum average $<N> ~\gg 100$ GeV, making it superweakly
coupled to matter ($g_A/\mu \sim 1/<N>$).

\end{document}